\documentclass[a4paper, 10pt, twocolumn]{article}
\usepackage[utf8]{inputenc}         % Schriftkodierung dieser Datei
\usepackage[english]{babel}          % für deutsche Dokumente

\usepackage{graphicx}               % optional für Grafiken
\usepackage{tabularx}               % optional für Tabellen
\usepackage{multirow}               % optional für Tabellen
\usepackage{url}                % optional für Internet Links

\usepackage[small,bf]{caption2}     % bitte für Bildunterschriften verwenden
\usepackage{parskip}
\usepackage{titlesec}
\usepackage{amsmath}    
\usepackage{hyperref}% optional für Formeln
\usepackage{booktabs}
\usepackage{pdflscape}

\titleformat{\section}{\normalfont\large\bfseries}{\thesection}{}{}
\titleformat{\subsection}{\normalfont\large\bfseries}{\thesection}{}{}
\titleformat{\paragraph}{\normalfont\bfseries}{\theparagraph}{}{}
\titlespacing{\section}{0pt}{6pt}{-1pt}
\titlespacing{\subsection}{0pt}{3pt}{-1pt}
\titlespacing{\paragraph}{0pt}{3pt}{-1pt}

\newcolumntype{Y}{>{\centering\arraybackslash}X}    %für Tabellen mit tabularx

\usepackage{tabu}

\begin{document}

\date{}                                         % kein Datum auf 1. Seite

\title{\vspace{-8mm}\textbf{\large
Dataset CYLinCF-01 creation pipeline: \\Circular cylinder in a cross flow\\Mach Number 0.03 and Reynolds Number 200  }}

% Hier die Namen und Daten der beteiligten Autoren eintragen
\author{
Stefan Schoder, Andreas Wurzinger  \\
\emph{\small TU Graz, Aeroakustik und Vibroakustik, IGTE, Inffeldgasse 18, A-8010 Graz, AT, Email: stefan.schoder@tugraz.at
} %\\
%$^2$ \emph{\small FAU Erlangen-Nürnberg, Aerodynamik und Akustik, LSTM, Cauerstraße 4, 91058 Erlangen, DE
%}
} \maketitle
\thispagestyle{empty}           % weder Kopf- noch Fußzeile auf Folgeseiten
% Beginn des eigentlichen Manuskripts

\section*{Abstract}

This article presents an aeroacoustic workflow (pipeline) to generate a flow and acoustic dataset for studying flow-induced sound in the context of a cylinder in cross flow. The numerical simulations are performed using OpenFOAM (from \href{www.openFoam.org}{openfoam.org} v9 or v10, or \href{www.openFoam.com}{openfoam.com} v2306) for the flow and openCFS for acoustics using the perturbed convective wave equation (PCWE) and the dilatation equation of Ribner. The workflow involves several steps, including the flow simulation, the acoustic simulation, and post-processing of the results in Python 3.9. The simulation steps are presented in all their details to verify your own segregated aeroacoustic solver. The analysis focuses on the acoustic characteristics of the flow, including sound pressure levels, frequency spectra, and directivity patterns. The results show good agreement with literature benchmarking datasets. The article concludes by discussing applications of the workflow for different cases that involve flow-induced sound generation.

\section*{Introduction}

Flow-induced noise is an omnipresent problem in a wide range of engineering applications, from aircraft and automotive design to environmental noise control. The noise generated by the flow around an object can lead to acoustic pollution. %Therefore, understanding and predicting flow-induced noise is essential for optimizing the performance of many industrial systems.
One of the most common geometries for studying flow-induced noise is a cylinder in cross flow \cite{margnat2023cylinder}. This simple configuration is relevant for many applications, such as heat exchangers, cooling towers, and offshore structures. The flow around the cylinder generates a wake, which interacts with the cylinder and leads to the emission of acoustic waves.

To support the development of numerical simulations techniques of flow-induced noise predictions, this benchmark dataset generation pipeline is published to the community. Since the simulation of both the flow and the acoustics is a challenging task, this dataset pipeline assures high quality from the beginning of your aeroacoustic developments. The coupling between the flow and the acoustics is described by the perturbed convective wave equation (PCWE) \cite{schoder2021aeroacoustic} for example, but can be modeled with any other aeroacoustic equation.

The flow is computed using the open-source software packages \textit{OpenFOAM} (v9 or v10 form openFoam.org or v2306 from OpenFoam.com) and the acoustic is computed using openCFS \cite{CFS}. These packages offer all methods required for an accurate flow acoustics simulation. Moreover, open-source software promotes reproducibility and collaboration, as the codes and data are freely available to the research community. You can use this example as a first verification of your segregated aeroacoustic solver. 
This article presents a hybrid aeroacoustic workflow \cite{schoder2019hybrid} for studying flow-induced noise in a cylinder in cross flow. The methodology is validated using data from the literature. We would like to encourage contributors having obtained nice visualizations or computations to share their knowledge with us.

% The first step of the workflow is ...

% The second step is ...

% The third step ...

% Finally, ... 

%The article is organized as follows. Section 2 presents the theoretical background for flow-induced noise and the numerical methods for simulating flow and acoustics. Section 3 describes the Aeroacoustic workflow for studying flow-induced noise in a cylinder in crossflow. Section 4 presents the validation of the methodology using experimental data and benchmark cases. Section 5 discusses the results and their implications for industrial applications. Finally, Section 6 summarizes the main contributions of the article and outlines directions for future research.

\section*{Workflow in short}

\subsection*{Flow Simulation}

The flow simulation is performed using OpenFOAM v9 or v10 (for v10 the transportProperties file must be renamed into physicalProperties, for versions of openFoam.com (e.g. v2306) the name stays transportProperties), a popular open-source computational fluid dynamics software package. The Navier-Stokes equations are solved using a finite-volume method, and the cylinder is modeled as a quasi-two-dimensional object with a structured mesh around it. The inlet boundary conditions is prescribed by the free stream velocity, at the outlet the pressure is set to zero and the cylinder surface is modeled as a no-slip wall. The pressure and velocity fields are computed, stored, and the results can be post-processed to obtain information about the flow characteristics, such as the wake structure and the drag coefficient.

\subsection*{Source Term Computation}
After the flow computation, the aeroacoustic sources are computed and interpolated onto the acoustic mesh (for details, see \cite{schoder2021application}).

\subsection*{Acoustic Simulation}
The acoustic simulation is performed using openCFS, an open-source software package for simulating multi-physical processes including acoustics. The linearized wave equation is used to model the acoustic waves, and the sound pressure levels and spectra are computed. The boundary conditions are set to be a sound-hard cylinder wall and free-field conditions otherwise.

\section*{Case description}

The first validation example demonstrates the impact of the different interpolation strategies on the near-field and far-field of acoustic sound propagation. We choose the 2D, laminar flow over a cylinder ($\mathrm{Re} = 200$, $\mathrm{Ma} =U0/c=0.029$, see Fig. 9) to report the impact. An unsteady flow field is observed, with a periodic vortex shedding at a Strouhal number of 0.205 and 0.41 \cite{zdravkovich1997flow}. This periodic fluid instability inside the wake of this cylinder generates tonal components in the sound field. The CFD was computed employing OpenFoam. The incompressible pressure was used to compute the aeroacoustic sources based on the PCWE. Afterward, this source term was conservatively interpolated to the acoustic mesh. Having obtained the sources, the acoustic propagation was computed. %https://www.worldscientific.com/doi/epdf/10.1142/S2591728520500322 We compare four acoustic source domain meshes. Table 5 sums up theinvestigated meshes ranging from refined acoustics meshes to coarse acoustic meshes.

%24 M. M. Zdravkovich,Flow Around Circular Cylinders: Volume 2: Applications(Oxford UniversityPress, 1997).

\section*{Generate a dataset using the pipeline}
%# Cylinder in a Crossflow
\subsection*{Description}
This problem considers a 2D cylindrical obstacle (diameter $d$=0.02 m) within an incompressible flow.
Inflow velocity $U_0$ = 10 m/s (Mach 0.029), kinematic viscosity $\nu$ = 0.001 m$^2$/s, Reynolds number $\mathrm{Re} \approx$ 200.
\begin{figure}[ht!]
    \centering
    \includegraphics[scale=0.3]{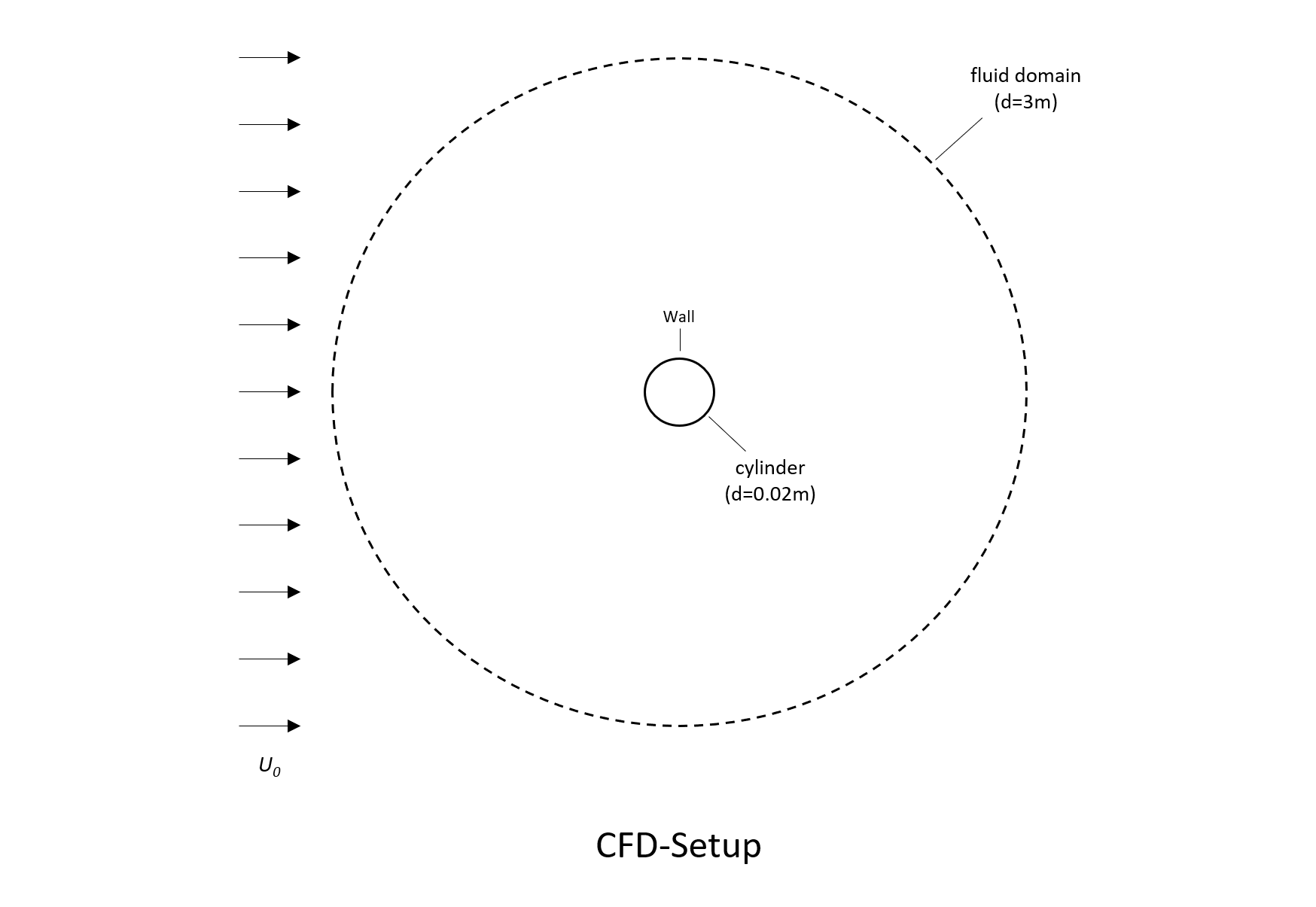}
    \caption{CFD Setup.}
    \label{fig:cfd}
\end{figure}
%![CFD Setup](cfd.png)
Air at 20$^\circ$C should be defined as material for the simulation (Material parameters are inside the \textit{'air.xml'} file.). The speed of sound of air can be assumed to be $c_0 \approx 343.5 \frac{m}{s}$. The computational aeroacoustic domain is depicted below.
\begin{figure}[ht!]
    \centering
    \includegraphics[scale=0.3]{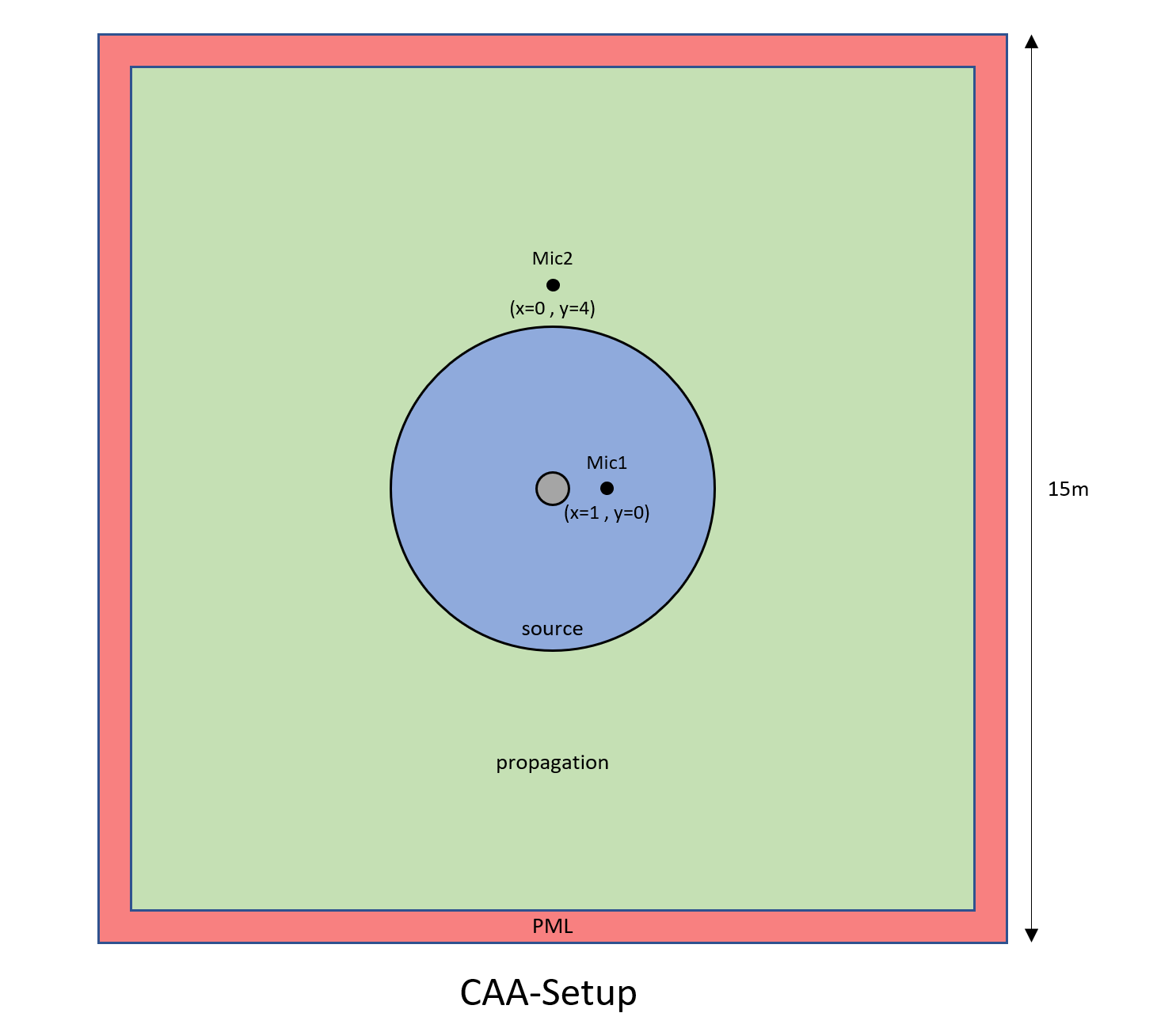}
    \caption{CAA Setup.}
    \label{fig:caa}
\end{figure}
%![CAA Setup](caa.png)
Inside the source region, a transient incompressible CFD Simulation is performed. Thereafter, aeroacoustic source terms can be computed from the obtained flow field. We then solve the Perturbed Convective Wave Equation (PCWE)
$$
\frac{1}{c^2} \, \, \frac{\mathrm{D}^2\psi^{\rm a}}{\mathrm{D} t^2} - \nabla \cdot \nabla \psi^{\rm a} = - \frac{1}{\bar \rho c^2}\, \frac{\mathrm{D} p^{\rm ic}}{\mathrm{D} t}
$$
for this low Mach number $\mathrm{Ma} \approx \frac{U_0}{c_0} = 0.03 \ll 0.3$ flow. The convective wave equation fully describes acoustic sound generated by incompressible flow structures and its wave propagation.
The only unknown is the acoustic scalar potential $\psi^{\rm a}$ . In order to receive the acoustic pressure, we have to derive the scalar potential and scale it with the density 
$$
p^\mathrm{a} = \rho_0 \frac{\mathrm{D} \psi^\mathrm{a}}{\mathrm{D} t} \,.
$$
For this problem, the convective part of the operator $\frac{\mathrm{D}}{\mathrm{D} t} = \frac{\partial}{\partial t} + \boldsymbol{\bar{v}} \cdot \nabla$ has a negligible effect on the acoustic field. Therefore, the convective part of the RHS source term $\boldsymbol{v} \cdot \nabla p^{\rm ic}$ is neglected for simplicity.

Here you can download the full simulation setup \href {https://opencfs.gitlab.io/userdocu/Applications/Singlefield/Acoustics/CylinderFlow/CylinderCrossFlow.zip}{CylinderCrossFlow.zip}. The convective part is considered in the \textit{'file\_PCWE'} files and neglected in the \textit{'file\_dpdt'} files.

%---------------------------
\subsection*{Perform Transient Flow Simulation with openFOAM}
%### Perform Transient Flow Simulation with openFOAM
Based on the Strouhal number for flows around obstacles of $\mathrm{St} \approx 0.2$ we estimate a vortex shedding frequency of
$$ 
f = \mathrm{St} \frac{U_0}{d} \approx 100 \,\mathrm{Hz}
$$
Therefore, we choose a time step of $\Delta t_\mathrm{CA} = 5 \cdot 10^{-4} \,\mathrm{s} \approx \frac{1}{20 f}$.
\begin{itemize}
    \item Set up the OpenFoam Mesh
    \item Define Boundary Conditions
    \begin{itemize}
        \item Velocity inlet
        \item Pressure outlet
    \end{itemize}
    \item Specify transient controls and solution methods
    \begin{itemize}
        \item Time stepping based on $\mathrm{CFL} < 1$: $\Delta t_{\mathrm{CFD}} = 5\cdot 10^{-5} \,\mathrm{s}$
        \item Standard solution methods
        \item Incompressible transient flow solver: \textit{'icoFoam'} using the PISO algorithm
    \end{itemize}    
    \item Run simulation until steady oscillation is set up
    \item Export pressure data as \textbf{EnSight Gold Case} (*.case) \textit{'foamToEnsight -fields '( "p.*" )' -time 0.5:0.7'}
    \begin{itemize}
        \item Standard CFD-Result file format
        \item Can be exported with OpenFoam, StarCCM++, Ansys Fluent, etc.
        \item Can be visualized with ParaView
    \end{itemize}    
\end{itemize}

%TODO flow field evaluation:
% Drag, Ligt
% Shading frequency
% Flow detachment point
% ... more ...

\subsection*{Interpolate the Flow Field and Compute Acoustic Sources}
Based on the Perturbed Convective Wave Equation (PCWE) the RHS source needs to be calculated. For this example, the convective part of the RHS source is neglected and only $\frac{\partial p^\mathrm{ic}}{\partial t}$ needs to be calculated.
\begin{itemize}
    \item Define data input
    \item Define partial time derivative
    \item Define interpolation filter
    \item Define data output
    \item Calculate interpolated acoustic sources: \textit{'cfsdat -p interpolation.xml -tX interpolation'} using \textbf{X} threads in parallel.
\end{itemize}

\subsection*{Acoustic Simulation Setup}
%### 
Finally, the acoustic propagation simulation is performed using the (already integrated) RHS values of 
$$
- \frac{1}{\rho_0 c_0^2} \int_\Omega \frac{\partial p^{\rm ic}}{\partial t}  \, \, d \Omega \, .
$$

\begin{itemize}
    \item Define I/O files
    \begin{itemize}
        \item Define data and mesh input
        \item Include material file (`air.xml`)
        \item Define output file formats
    \end{itemize}  
    \item Define computational domain
    \begin{itemize}
        \item Assign material to regions (air at 20 degrees)
        \item Define non-conforming interfaces between the different regions/parts (Nitsche-type Mortaring)
        \item Define microphone points for evaluation 
    \end{itemize}  
    \item Define simulation
    \begin{itemize}
        \item Specify simulation type (transient)
        \item Define the acoustic PDE (Potential formulation)
        \begin{itemize}
            \item Specify regions to use in calculation
            \item Specify used NC interfaces
            \item Specify damping for perfectly-matched-layer (PML) regions
            \item Specify background flow for convective wave operator
            \item Define RHS node values (multiply with $- \frac{1}{\rho_0 c_0^2}$)
            \item Define output results
        \end{itemize}
    \end{itemize} 
    \item Run the acoustic propagation simulation: \textit{'cfs -p propagation.xml -tX propagation'} using \textbf{X} threads in parallel.
\end{itemize}

\subsection*{Post-Processing Simulation Results}
View the acoustic field result with ParaView nightly or 5.12 and greater versions that support the native CFS format directly. In addition, the results can be post-processed using the shipped python 3.9 scripts that evaluate different properties of the acoustics, such as the directivity. More on the post-processing can be found in the openCFS documentation on opencfs.org.

\section*{Selected Applications using openCFS, hybrid Aeroacoustics and Benchmarking Datasets}
Further contributions using workflows implemented with openCFS \cite{CFS} and \textit{openCFS-Data} \cite{CFSDAT,schoder2023offline} are presented in this section. For instance, the software packages were used to implement the stochastic noise generation and radiation (SNGR) method based on an incompressible RANS simulation \cite{schoder2023opencfs} and applied to cavity noise \cite{weitz2019approach,weitz2019numerical}. Furthermore, the cavity noise simulations of the SNGR method were compared to hybrid aeroacoustic workflow simulations using high-resolved flow data was also used to compute the cavity noise, recently\cite{schoder2018aeroacoustic,schoder2019hybrid,schoder2020conservative}. The spatial derivatives of source terms were computed by a radial basis function scheme \cite{schoder2020radial,schoder2020aeroacoustic}. Recently developed equations can be leveraged by the implemented methods presented in the work described in \cite{schoder2022aeroacoustic,schoder2022cpcwe,schoder2023acoustic}. Also, the combination with a Helmholtz decomposition  \cite{schoder2020postprocessing,schoder2020postprocessing2,schoder2019helmholtz,schoder2022post,schoder2022helmholtz} is possible to establish a convective scalar wave equation according to the APE theory. Furthermore, the software and methodologies developed can be valuable for automotive OEMs sound prediction workflows for the passenger cabin \cite{engelmann2020generic,weitz2019numerical}, combustion auxiliary parts \cite{freidhager2021simulationen} and the HVAC systems \cite{schoder2020numerical,maurerlehner2022aeroacoustic}. The hybrid aeroacoustic workflow was found to be applicable for low-pressure axial fan noise computations (e.g., heat pump ventilators or cooling fans in the automotive industry) \cite{schoder2020computational,tautz2018source,schoder2021application,tieghi2022machine,tieghi2023machine,schoder2022dataset}, high-lift fans for air-taxis \cite{schoder2023affordable}, radial fans (currently under development \cite{uffinger2023development}), the noise propagation simulations of the turbocharger compressor \cite{freidhager2022applicability,kaltenbacher2020modelling,freidhager2020influences}, the acoustics of fluid-structure-acoustic-interaction processes \cite{schoder2020hybrid,schoder2022pcwe} with special focus on human phonation \cite{valavsek2019application,zorner2016flow,schoder2021aeroacoustic,falk20213d,lasota2021impact,maurerlehner2021efficient,schoder2022learning,lasota2023anisotropic,kraxberger2022machine,schoder2023implementation,kraxberger2023alignment}. Potential nonphysical behavior generated by the source computation can be identified using the methods in \cite{schoder2022error} to assess statistical convergence. Finally, the sound sources of an PIV-measured flow field were evaluated in \cite{nager2023investigation} for a wide range of operating conditions to learn about the FSAI process in human phonation.

\section*{Summary and Benchmark Proposal}

In this article, a hybrid aeroacoustic workflow for studying flow-induced noise of a cylinder in cross flow using OpenFOAM for the flow and openCFS for the acoustics has been presented. The methodology has been validated using data from the literature. The proposed data generation pipeline based on the PCWE can be used to test your aeroacoustic workflows and new equations in aeroacoustics. This is a community call, if you generate data with this workflow and want to discuss the results or detect discrepancies in our methods used (we were not aware of), please report them to us.

\bibliographystyle{abbrv}
\bibliography{references} 

\end{document}